%Paper: hep-th/9301131
%From: DEPIREUX@LPS.UMONTREAL.CA
%Date: Sun, 31 Jan 1993 14:57:47 -0500 (EST)

%%%%%%%%%%%%%%%%%%%%%%%%%%%%%%%%%%%%%%%%%%%%%%%%%%%%%%%%%%%%%%%%%%%%%%
%%%%%%%%%%%%%%                                          %%%%%%%%%%%%%%
%%%%%%%%%  Integrable Supersymmetry Breaking Perturbations   %%%%%%%%%
%%%%%%%%%                      of                            %%%%%%%%%
%%%%%%%%%        N=1,2 Superconformal Minimal Models         %%%%%%%%%
%%%%%%%%%                                                    %%%%%%%%%
%%%%%%%%%                                                    %%%%%%%%%
%%%%%%%%%     by Didier A Depireux and Pierre Mathieu        %%%%%%%%%
%%%%%      This is a Tex file, using the macro  harvmac.tex      %%%%%
%%                                                                  %%
%%%%%%%%%%%%%%%%%%%%%%%%%%%%%%%%%%%%%%%%%%%%%%%%%%%%%%%%%%%%%%%%%%%%%%

\input harvmac.tex

\overfullrule=0pt
\Title{\vbox{\baselineskip12pt
\hbox{LAVAL PHY-20-93}
\hbox{hep-th@xxx/9301131}}}
{\vbox{
\centerline{ Integrable Supersymmetry Breaking Perturbations}\vskip.5cm
\centerline{ of }\vskip.5cm
\centerline{ N=1,2 Superconformal Minimal Models}}} \vskip1cm
\centerline{Didier A.~Depireux and Pierre Mathieu}
\smallskip
\centerline{D\'epartement de Physique,}
\centerline{Universit\'e Laval,}
\centerline{Qu\'ebec, Canada G1K 7P4.}
\vskip .3in
{\bf Abstract:}  We display a new integrable perturbation for both N=1 and N=2
superconformal minimal models. These perturbations break supersymmetry
explicitly. Their existence was expected on the basis of the classification of
integrable perturbations of conformal field theories in terms of distinct
classical KdV type hierarchies sharing a common second Hamiltonian structure.
%related to the conformal algebra.
\vskip .3in

\Date{ Jan.1992.}

%Macros
%%%%%%%%%%%%%%%%%%%%%%%%%%%%%%%%%%%%%%%%%%%%%%%%%%%%%%%%%%%%%%%
%%%%%%%%%%%%%%%%%%%DEFINITIONS%%%%%%%%%%%%%%%%%%%%%%%%%%%%%%%%%
%
\def\frac#1#2{{#1 \over #2}}
\def\ha{{1 \over 2}}
\def\pa{\partial}
\def\vac{\mid 0 >}
%

% References
\lref\un{A.B.Zamolodchikov, JETP Lett. 46 (1987) 160; Int.J.Mod.Phys. A3
(1988) 4235; A4 (1988) 743; Adv. Pur. Math. 19 (1989)641.}
\lref\trois{B.A.Kupershmidt and P.Mathieu, Phys.Lett. B227 (1989) 245.}
\lref\quatre{P.Mathieu, Nucl.Phys. B336 (1990) 338.}
\lref\sept{T.Eguchi and S.K.Yang, Phys.Lett. B 224 (1989) 373.}
\lref\huit{T.J.Hollowood and P.Mansfield, Phys.Lett B226 (1989) 73.}
\lref\MatWalt{P.Mathieu and M.A.Walton, Phys.Lett. B254 (1991) 106.}
\lref\onze{C.-A.Laberge and P.Mathieu, Phys.Lett. B215 (1988) 718.}
\lref\douze{P.Labelle and P.Mathieu, J.Math.Phys. 32 (1991) 923.}
\lref\treize{P.Fendley, S.D.Mathur, C.Vafa and N.Warner, Phys.Lett. B243
(1990) 257; P. Fendley, W. Lerche, S.D. Mathur and N. P. Warner, Nucl.
Phys. B348 (1991) 66.}
\lref\PierreJMP{P.Mathieu, J.Math.Phys. 29 (1988) 2499.}
\lref\PierrePLA{P.Mathieu, Phys.Lett. A128 (1988) 169.}
\lref\PierrePLB{P.Mathieu, Phys.Lett. B203 (1988) 287.}
\lref\Hlavaty{L.Hlavaty, Phys.Lett. A137 (1989) 73.}
\lref\Zhang{D.-G.Zhang and B.-Z.Li, Phys.Lett A171 (1992) 43.}
\lref\InamiKanno{T.Inami and H.Kanno, Commun.Math.Phys, 136 (1991)
519.}
\lref\KuperKdV{B.A.Kupershmidt, Phys.Lett. A108 (1984) 213.}
\lref\WZW{L.Bonora, Y.-Z.Zhang and M.Martellini, Int.J.Mod.Phys. A6 (1991)
1617.}
\lref\Jeremy{J.Schiff, ``\it Self-dual Yang-Mills and the Hamiltonian
Structures of Integrable Systems\rm'', hep-th/9211070.}
\lref\thiel{K. Thielemans, Int. J. Mod Phys. C 2 (1991) 787.}

\secno=-1
\newsec{Posing the problem}

In two dimensional quantum field theory, integrability singles out the class
of tractable models. These can be efficiently formulated as integrable
perturbations of some conformal field theory\un. There exist few guiding
principles which can be used to classify the full set of integrable
perturbations of a given (extended) conformal field theory, but the most
powerful and universal one appears to be the following: the number of
integrable perturbations is given by the number of integrable hierarchies of
the KdV type, whose second Hamiltonian structure is associated to the extended
conformal algebra, and which have distinct first Hamiltonian
structures\trois\quatre\MatWalt. A  one-to-one correspondence between
perturbating fields and the KdV hierarchies can be obtained via the
associated Toda systems\sept\huit. In the Feigin-Fuchs representation, the
perturbating field is represented by the part of the Toda Hamiltonian which
is not a screening operator.

For the usual Virasoro minimal models, there are three integrable perturbations
($\phi_{1,3}$, $\phi_{1,2}$ and $\phi_{2,1}$), corresponding to the existence
of three integrable hierarchies sharing the second Poisson structure of the KdV
equation but having distinct Poisson brackets for the first Hamiltonian
structure. These are the KdV hierarchy itself and the two reductions of the
Boussinesq hierarchy\trois. Their Toda system is related to the affine $su(2)$
and twisted $su(3)$ algebras respectively (the asymmetry of the later giving
rise to two KdV type hierarchies). For the N=1 superconformal minimal models,
the only supersymmetric integrable perturbation is $\hat\phi_{1,3}$ \quatre
(the hat denotes a superfield). It corresponds to the unique (space)
supersymmetric KdV type system whose second Hamiltonian structure is the
classical form of the N=1 superconformal algebra\PierreJMP. The underlying
affine algebra is twisted $osp(2,2)$\InamiKanno. In \MatWalt, this approach was
used to predict the existence of three supersymmetric integrable perturbations
for the N=2 minimal models (also found in \treize), given that there are three
integrable N=2 (space) supersymmetric KdV hierarchies\onze\douze. The
corresponding perturbating fields are the chiral superfields $\Phi^l$ with
$l = 1,2,K$, where $K$  is related to the central charge by
$K = 2c/(3-c).$\footnote{${}^1$}{These ideas have also been extended to
parafermionic models via the non-linear Schr\"odinger equation in disguised
form \Jeremy.}.

However, it is known that there exists one\footnote{${}^2$}{See the appendix
about the {\it one}.} integrable fermionic (but not supersymmetric) extension
of the KdV equation whose second Hamiltonian structure is the N=1
superconformal algebra. It is the ``super KdV'' equation of
Kupershmidt\KuperKdV. Since this equation is not actually supersymmetric, we
call it the KuperKdV equation. Its underlying algebraic structure is
$osp(1,2)$. Furthermore, its $o(2)$ integrable extension (connected to
$osp(2,2)$) turns out to be related to the N=2 superconformal algebra\onze.
Hence, according to the above organizing principle, one expects these
hierarchies to be related to supersymmetry breaking integrable perturbations
of the N=1,2 superconformal minimal models. Here we show that this is indeed
the case.

Without using the Feigin-Fuchs representation, it is possible to guess
which perturbating field is associated to each hierarchy: the most natural
relevant supersymmetry breaking perturbation is simply the lowest component of
the superfield whose top component yields the integrable supersymmetry
preserving perturbation. To be more precise, we consider an N=1 superfield
\eqn\la{
\hat\phi = \phi + \theta \psi~. }
The supersymmetric transformations of the component fields are
\eqn\laa{
\delta \phi = \eta\psi \qquad \delta \psi = \eta \del \phi}
where $\eta$ is a constant anticommuting parameter.
The superintegral $\int\, dz\, d\theta \hat\phi$ is just $\int\, dz \psi,$
which is manifestly supersymmetric. However, $\int\,  dz \phi$ is
 not supersymmetric invariant. $\phi$ is referred to as the lower
component of the superfield. If $\psi$ is relevant (with scaling dimensions
smaller than two), then so is $\phi.$ Indeed, perturbing the N=1
superconformal minimal models with the lower component of $\hat\phi_{1,3}$
leads to a (presumably) infinite sequence of conservation laws whose
classical limit agrees with the KuperKdV conserved integrals.

In the N=2 case, three choices are possible. But again, a naturalness criterion
would select the lower component of $\Phi^K$ as being the correct choice.
Indeed, $\Phi^K$ is the N=2 analog of $\phi_{1,3}$ (or $\hat \phi_{1,3}$), and
in any conformal field theory, the appropriate generalization of  $\phi_{1,3}$
is an integrable perturbation. This makes the $\Phi^K$ perturbation `more
fundamental' than the other two. As a matter of fact, the massive theories
obtained by perturbing the $N=2$ minimal models with the lower component of
$\Phi^K$ appear to be integrable and their conservation laws are exactly the
quantum generalization of the $o(2)$ KuperKdV ones.

When treating a
problem in which supersymmetry is not preserved, one needs to work out
everything in terms of components. In such a case, the presence of the $u$(1)
symmetry induces a little catch. It appears at first sight that perturbing
N=2 superconformal models with one component or the other of any chiral
primary field (they are all relevant) produces non-trivial conservation laws
(even without using the degeneracy equations of the perturbating field).
These conservation laws all happen to be expressible solely in terms of a
twisted energy-momentum tensor $\tilde T$. Actually, a closer look shows
that switching on the perturbation does not induce any $\bar z$ dependence on
$\tilde T$, which suggests that the critical point has not been left. In
fact, this is exactly what happens. With respect to $\tilde T$, all
perturbating fields become marginal. This simple looking observation, when
transposed in the context of perturbed $\hat{su}$(2)$_k$ models, accounts for
most of the conservation laws found in \WZW (for a perturbation with an
arbitrary primary field). We will report on this problem elsewhere.

\newsec{N=1}

The procedure for computing integrals of motion in perturbed conformal field
theory (to first order) is by now rather standard. In terms of the operators
\eqnn\lb
$$
\eqalignno{
\Gamma_n  & = {1\over 2 i \pi} \oint\,dz z^n \phi_{1,3}(z) \quad,\cr
\Lambda_n & = {1\over 2 i \pi} \oint\,dz z^{n+1/2} \psi_{1,3}(z)
\quad,\qquad &\lb  \cr}$$
where $\phi_{1,3}$ and $\psi_{1,3}$ are the components of the superfield
$\hat\phi_{1,3}$ (cf eq. \la), $F_s$, the conserved quantity of spin $s$,
(a differential polynomial in $T(z) = L_{-2} I$ and $G(z) = G_{-\frac32} I$)
is characterized by the fact that $\Gamma_0 F_s$ is a total derivative.
To show this, it is necessary to use the degeneracy equation of the perturbing
field,
whose component form reads \quatre
\eqnn\lc
$$
\eqalignno{
L_{-1} \psi(0) \vac & = \ha (2h+1) G_{-\frac32} \phi(0) \vac \quad,\cr
L_{-1}{}^2 \phi(0) \vac & = (2h+1) [L_{-2} - \ha G_{-\frac32} ] \phi(0) \vac
 \quad,\qquad &\lc  \cr}$$
with $(\psi,\phi,h) = (\psi_{1,3},\phi_{1,3},h_{1,3}).$ With the central
charge parametrized as
\eqn\ld{
c = \frac 32 (1-\frac 8{p(p+2)}) ~, }
$h_{1,3}$ reads
\eqn\le{
h_{1,3} = \frac{(p-2)}{2(p+2)}~. }
The vectors associated with the first few conserved densities are
\eqnn\lg
$$
\eqalignno{
F_2 & = L_{-2} \vac \quad,\cr
F_4 & = (L_{-2}{}^2 - 2 \frac{h-1}{2h+1} G_{-\frac32} G_{-\frac52})
\vac\quad,\cr
F_6 & = (L_{-2}{}^3 + 6 \frac{h-1}{2h+3} L_{-2} G_{-\frac52} G_{-\frac32} + 8
\frac{(h-1)^3}{(2h+1)(2h+3)} G_{-\frac72} G_{-\frac52} \cr
&\qquad + (\frac{4h^3 + 4h^2 - 31 h + 8}{4(2h+1)(2h+3)} ) L_{-3}{}^2 )
\vac\quad, \cr
F_8 & = (L_{-2}{}^4 + a_1 L_{-3}{}^2 L_{-2} + a_2 L_{-4}{}^2 + a_3 G_{-\frac92}
G_{-\frac32} + a_4 G_{-\frac72} G_{-\frac52} \cr
&\qquad + a_5 L_{-2}{}^2
G_{-\frac52} G_{-\frac32} + a_5 L_{-4} G_{-\frac52} G_{-\frac32}) \vac
\quad\cr}$$
 where
$$\eqalignno{
a_1 & = \frac{4h^3 - 4 h^2 - 67 h + 4}{(2h+1)(2h+5)}\quad,\cr
a_2 & = \frac{24h^5 + 28 h^4 - 1386 h^3 + 1713 h^2 + 215 h + 225}{15
(2h+1)^2(2h+5)} \quad,\cr
a_3 & = \frac{24 (h-4)(h-1)^2(16 h^2 - 6 h + 13)}{5(2h+1)^2(2h+5)} \quad,\cr
a_4 & = \frac{16(h-1)^2(2h-5)}{(2h+1)(2h+5)}\quad,\cr
a_5 & = \frac{12 (h-1)}{2h+5}\quad,\cr
a_6 & = \frac{4 (h-1) (-6h^2 + 19 h + 5)}{(2h+1)(2h+5)}
\quad.\qquad &\lg  \cr}$$
We have checked their commutativity  (using the
Mathematica package of \thiel).
With the rescalings
\eqn\lh{
T = - \frac{c}6 u \quad,\qquad G = \frac{c}3 \xi \quad, }
these conserved quantities can be checked to reduce to those
of the KuperKdV equations \KuperKdV
\eqnn\lhh
$$
\eqalignno{
u_t & = - u_{xxx} + 6 u u_x + 3 \xi \xi_{xx} \quad, \cr
\xi_t & = - 4 \xi_{xxx} + 6 \xi_x u + 3 \xi u_x
\quad,\qquad &\lhh  \cr}$$
in the limit of $c \rightarrow \infty$ (which can
be realized by $p \rightarrow -2$ or $h \rightarrow - \infty$). The first
few conserved densities for the system \lhh\ are
\eqnn\lhi
$$
\eqalignno{
h_2  = & u \quad, \quad\cr
h_4  = & u^2 - 4 \xi \xi_{xx} \quad, \quad\cr
h_6  = & u^3 + \ha u u_x + 12 u \xi_x \xi + 8 \xi_{xx} \xi_x \quad, \quad\cr
h_8  = & u^4 + 2 u_x^2 u + \frac{1}5 u_{xx}^2 + 24 u^2 \xi_x \xi
- 24 u_{xx} \xi_x \xi + \cr
& \quad 32 u \xi_{xx} \xi_x + \frac{64}5 \xi_{xxx} \xi_{xx}
\quad.\qquad &\lhi  \cr}$$
On the other hand, no conserved densities with odd spins have been found and
these do not exist even in the classical case.

\newsec{N=2}

An (anti)-chiral superfield $\tilde\Phi$
satisfies the constraints $D^- \tilde\Phi =0 $, which implies a component
expansion of the form
\eqn\li{
\tilde\Phi = \varphi + \ha \theta^- \psi^+ - \theta^+\theta^-
\del \varphi ~. }
Let $\tilde\Phi$ stand for $\Phi^K$ where $K$ refers to the following
parametrization of the central charge
\eqn\lj{
c = 3 ( 1 - \frac2{K+2} ) ~.}
The conformal dimension and the $u(1)$ charge of $\tilde\Phi$ are given by
\eqn\lk{
h = - q = \frac{K}{2(K+2)} }
(which fixes our relative normalization for the $u(1)$ current) while its
degeneracy equation in component form reads
\eqnn\ll
$$
\eqalignno{
( & L_{-1} + 2 J{_1} ) \varphi(0) \vac = 0  \quad,\cr
( & L_{-1} + 2 J{_1} ) \psi^+ \ (0) \vac = - 2 G_{-\frac32}^+ \varphi(0)
\vac  \quad,\cr
{} & G_{-\frac32}^- \varphi(0) \vac = 0  \quad,\cr
( & L_{-1}{}^2 + 2 J_{-1} - 2 L_{-2}) \varphi(0) \vac = \ha
G_{-\frac32}^- \psi^+(0) \vac
\quad.\qquad &\ll  \cr}$$
We consider the perturbation $\int\,dz~ \varphi$ (ignoring the
antiholomorphic part). For this we introduce the quantities
\eqnn\lm
$$
\eqalignno{
\Gamma_n    & = {1\over 2 i \pi} \oint\,dz z^n \varphi(z) \quad,\cr
\Lambda_n^+ & = {1\over 2 i \pi} \oint\,dz z^{n+1/2} \psi^+(z)
\quad,\qquad &\lm  \cr}$$
whose commutators with the generators of the $N=2$ superconformal algebra are
\eqnn\ln
$$
\eqalignno{
[\Gamma_m,L_n]   & = [-h(n+1) +(m+n+1)] \Gamma_{m+n} \quad,\cr
[\Gamma_m,G_n^+] & = \ha \Lambda_{m+n}^+  \quad,\cr
[\Gamma_m,G_n^-] & = 0                    \quad,\cr
[\Gamma_m,J_n]   & = h \Gamma_{m+n} \quad,\cr
[\Lambda_m^+,L_n]  & = [- (n+1)(h+\ha) + (m+n+\frac32) ]
\Lambda_{m+n}^+ \quad,\cr
\{ \Lambda_m^+,G_n^+ \} & = 0
\quad,\cr \{ \Lambda_m^+,G_n^- \} & = [ - 4 h (2n+1) +4(m+n+1)] \Gamma_{m+n}
\quad,\cr [\Lambda_m^+,J_n]   & = (h-\ha) \Lambda_{m+n}^+
\quad.\qquad &\ln  \cr}$$
Notice that $\Lambda_{m+n}^+$ is odd and that
\eqn\lo{
\Gamma_{-m-1} I = \frac1{m!} L_{-1}{}^m \varphi \quad,\quad
\Lambda_{-m-\frac32}^+ I = \frac1{m!} L_{-1}{}^m \psi^+~. }
$I$ being the identity field.
Proceeding as for $N=1$, the first few conserved densities with integer
spins are found to be
\eqnn\lp
$$
\eqalignno{
\tilde F_2 = & L_{-2} \vac \quad,\cr
\tilde F_3 = & (L_{-2} J_{-1} - \frac2{3h} J_{-1}{}^3 + \frac{h}2
G_{-\frac32}^- G_{-\frac32}^+ ) \vac \quad, \cr
\tilde F_4 = & (L_{-2}{}^2 + a_1 L_{-3} J_{-1} + a_2 J_{-2}{}^2 + a_3 L_{-2}
J_{-1}{}^2 + a_4 J_{-1}{}^4 \cr
&\quad + a_5 G_{-\frac52}^- G_{-\frac32}^+ + a_6
G_{-\frac32}^- G_{-\frac32}^+ J_{-1} ) \vac \quad, \cr
{\rm where} & {} \cr
a_5 & - a_6 (h-1) = 0 \quad , \cr
- 2 & (h-1) + a_1 h - a_3 h (h-1) + a_6 (2h-1) = 0 \quad , \cr
a_1 & (h-1) - 2 a_2 h + a_3 (h-1)^2 + 2 a_4 h (h^2 - 3 h + 1) = 0
\quad.\qquad &\lp  \cr}$$
At this point, we see that there are three arbitrary parameters in $\tilde
F_4.$ This unusual feature will be commented on shortly\footnote{${}^3$}{A
similar situation is observed for the perturbation $\int~dz \psi^+$ treated in
terms of components. But in this case, we can rely on supersymmetry to fix all
undetermined coefficients. In this way, the ambiguity discussed below is
bypassed.}. Nevertheless, $\tilde F_3$
is sufficient to make an unambiguous contact, in the classical limit, with
the  $o(2)$ KuperKdV equation. This equation can be written exactly like the
usual KuperKdV equation but with the replacement \onze\
\eqnn\lq
$$
\eqalignno{
&u \rightarrow q = u - w^2 \quad, \cr
&\xi \rightarrow \Psi = e^{\sigma(\pa^{-1} w)} \pmatrix{\xi_1\cr \xi_2\cr}
\quad,\quad \sigma = \pmatrix{0 & 1 \cr -1 & 0 \cr }
\quad.\qquad &\lq  \cr}$$
together with $w_t = 0 .$ To investigate the classical limit, we set
\eqn\lr{
T = - \frac{1}6 c u \quad,\quad G^{\pm} = \frac{i}3 c \xi^{\pm} \quad,\quad
J = - \frac{i}6 c w \quad,\quad}
where $\xi^{\pm} = (\xi_1 \pm i \xi_2)/\sqrt{2},$ and let $c = 6 h
\rightarrow \infty.$ $\tilde F_3$ reduces then to the product
$\xi_1 \xi_2$ (up to a multiplicative factor).
In terms of $\Psi$ it reads $\Psi^T\sigma\Psi$ ($T$ stands for transpose)
and this is easily checked to be conserved for the $o(2)$ KuperKdV equation.
\footnote{${}^4$}{This system admits conservation laws at all integer
degrees.   In the reduction $\xi_2=w=0$, even degree densities reduce to
those of the KuperKdV equation, while those at odd degrees vanish.}

Now this equation is somewhat `degenerate' in that the field associated to the
$u(1)$ current is time independent. This reflects itself in the fact that with
respect to the first Hamiltonian structure, the Poisson bracket of $w$ with
itself is zero. As a consequence, the recursive generation of the classical
conservation laws does not fix all the parameters \onze.  In particular, for
$\int \tilde h_4$, three parameters are left undetermined. Similarly, the
commutativity of $\int \tilde h_3$ and $\int \tilde h_4$ leaves two parameters
undetermined.

So apparently, this degeneracy extends to the quantum case, given that
the direct determination of $\tilde F_4$ contains three free
parameters.  Moreover, if we start from a generic form for $\tilde F_4$
and impose (using the package of \thiel)
\eqn\lz{[\int dz \tilde F_3,\int dz \tilde F_4]=0\quad ,\quad}
we find that four coefficients get determined exactly
\eqn\cof{
a_1 = - a_6 = {-4\over 2h-1},\quad a_3 = {-2(2h-3)\over h(2h-1)},\quad
a_5 = {4(h-1)\over 2h-1},\quad}
while $a_2$ and $a_4$ remain undetermined, precisely as in the
classical case.  The above values are compatible with the conditions
\lp\ and they reduce, in the classical limit, to the coefficients
of $ \tilde h_4$.  The undetermined coefficients can be fixed only from
the commutation with the higher order conservation laws.

In the perturbed theory, there are also conservation laws with
half-integer degrees, whose first few related vectors are
\eqnn\ls
$$
\eqalignno{
\tilde F_{\frac32} = & G_{-\frac32}^- \vac \quad,\cr
\tilde F_{\frac52} = & J_{-1} G_{-\frac32}^- \vac \quad, \cr
\tilde F_{\frac72} = & [J_{-1}{}^2 - h J_{-2} + a (L_{-2} -
\frac{h-1}h J_{-2})] G_{-\frac32}^- \vac  \quad,\qquad &\ls  \cr}$$
where $a$ is undetermined.  At first sight, this is somewhat surprising
because such conservation laws are not present at the classical level,
and this would be regarded as a strong indication that they should not
be there either in the quantum case.  However, a closer look shows that
they do not provide integrals of motion for the quantum $o(2)$ KuperKdV
equation, simply because they do not commute with $\int dz \tilde F_4$,
the defining Hamiltonian of the system when formulated canonically.
They do not commute either among themselves.  Their interest, if any, is
thus rather limited.

For other perturbations (in particular with the lowest component of the
chiral superfield $\Phi^l,$ $l=1,~2$) no conservation laws were found except
for a trivial sequence which we now discuss. From the commutators \ln, it
follows that
\eqn\lt{
[ \Gamma_0, L_{-n} - (n-1) \frac{h-1}{h} J_{-n} ] = 0 \quad.\quad}
This means that the perturbation commutes with the field
\eqn\lu{
\tilde T = T - \frac{h-1}{h} J'  \quad,\quad}
and any of its derivatives.
It thus implies that any differential polynomial in $\tilde T$
commutes with $\Gamma_0.$ This result only uses the fact that $\varphi$ is
the lower component of a chiral field but it is independent of the
degeneracy equation of the field under consideration. This implies that \lt\
holds for the perturbation by the lower component of {\it any} chiral
field\footnote{${}^5$}{For supersymmetry preserving perturbation, where
$\Gamma_0$ is replaced by
$\Lambda_{\ha}^+,$ the analog of \lt\ is
$$
[ \Lambda_{-\ha}^+, L_{-n} - (n-1) J_{-n} ] = 0 \quad.\quad $$
}.
However, with respect to $\tilde T$, the conformal dimension
of the lower component of any chiral primary fields is one, which means
that the perturbation is marginal. It thus acts as a simple
twist, a supersymmetry breaking term. But it does not drive the system
off-criticality.

\appendix{A}{}

The most general local fermionic extension of the KdV equation
is
\eqnn\lv
$$
\eqalignno{
u_t   & = - u_{xxx} + 6 u u_x - 3 \xi \xi_{xx} \quad,\quad\cr
\xi_t & = - a \xi_{xxx} + b u \xi_x + c u_x \xi \quad,\quad & \lv \cr}
$$
where $\xi$ is a fermionic field. The coefficients of the two
nonlinear terms in the first equation can of course be modified by
a rescaling the two fields.  However, they have been fixed in order to
exclude their possible vanishing, since in that case the system becomes
trivial.  The canonical formulation of the system in terms of the Poisson
structure which is the  classical limit of the superconformal algebra,
 forces the relations $c=3$ and $b=2+a.$ The supersymmetric KdV equation
corresponds to $a=1$ while $a=4$  for the Kupershmidt system. In \PierrePLB,
it was shown that the one parameter family of systems related to the
superconformal algebra, is integrable only for these two values of $a.$ The
same conclusion was obtained from the Painlev\'e analysis of the above more
general three parameters systems \PierrePLA. However, it has been argued in
\Hlavaty\ that the system $a=1$, $b=c=6$ is also Painlev\'e admissible (see
also \Zhang\ for further support on integrability). But this latter system is
trivial in the sense that the change of variables $ u = v + \xi (\del^{-1}
\xi)$ transforms the first equation into the usual KdV equation (i.e. the
fermionic field decouples) without affecting the second one.

\vskip.5cm

\vskip2truecm
{\bf Acknowledgements.}\par

We acknowledge useful discussions with D. Bernard, P. Di Francesco, N.
Sochen and M. Walton.  This work was supported by NSERC (Canada), FCAR
(Qu\'ebec), and BSR (Universit\'e Laval).

\vskip1truecm
\hrule
\vskip1truecm
\listrefs

\end